\documentstyle[preprint]{jpsj}
\input epsf

\newcommand{\lmk}{\left(}
\newcommand{\rmk}{\right)}

\newcommand{\lkk}{\left[}
\newcommand{\rkk}{\right]}

\newcommand{\beq}{\begin{equation}}
\newcommand{\eeq}{\end{equation}}
\newcommand{\beqa}{\begin{eqnarray}}
\newcommand{\eeqa}{\end{eqnarray}}
\newcommand{\lab}{\label}

\newcommand{\gtrsim} {~ \raisebox{-1ex}{$\stackrel{\textstyle >}{\sim}$} ~} 
\newcommand{\lesssim} {~ \raisebox{-1ex}{$\stackrel{\textstyle <}{\sim}$}~}
\begin{document}
%\if0
\def\runtitle{Probing the cosmic equation of state with a laser interfereometer}
\def\runauthor{Naoki {\sc Seto} and Jun'Ichi {\sc Yokoyama}}

\title{Probing the equation of  state of  the early universe with
a space laser interferometer}
\author{ Naoki {\sc Seto} and Jun'Ichi {\sc Yokoyama}${}^1$ }
\inst{
Theoretical Astrophysics, Mail code 130-33, California Institute of Technology, Pasadena,
CA 91125, USA\\
${}^1$Department of Earth and Space Science, Graduate School of
  Science, Osaka University, Toyonaka 560-0043\\

}

\abst{
We propose a method to probe the
equation of 
state of the early universe and its evolution, using  the stochastic
gravitational wave background from  inflation. A small deviation from
 purely  radiation dominated universe 
($w= 1/3$) 
would be clearly imprinted on the gravitational wave spectrum
 $\Omega_{GW}(f)$ due to the
 nearly scale 
 invariant nature of inflationary
generated waves.
}

%\pacs{PACS number(s): 95.55.Ym 04.80.Nn, 98.62.-g }
%\fi

\maketitle

%\baselineskip 8mm
%%%%%%%%%%%%%%%%%
%\section{Introduction}
%%%%%%%%%%%%%%%%5
Gravitational waves generated in the
 early universe are one of the most interesting
targets in 
observational cosmology.  Though  direct detection of  waves is not a
simple task due to  their extremely weak interaction, they could serve  as an
invaluable
fossil of the early universe that can be hardly probed by other
observational means \cite{Maggiore:2000vm,Allen}. 
The most realistic generation mechanism of the
stochastic gravitational wave
background in the early universe is  quantum effects 
during inflation  \cite{gwinf}. 
While  the fact that we live in a large and   nearly
 homogeneous universe alone is  sufficient to prove the existence
 of an inflationary epoch in the early universe \cite{oriinf}, 
recent observations of
 the WMAP satellite \cite{Bennett:2003bz} have provided a more sophisticated confirmation of
 inflation. This comes from that super-horizon-scale perturbations that
were 
 generated  in the very early universe present a negative
 correlation between temperature anisotropy and E-mode polarization
 (TE) \cite{Peiris:2003ff}.  Note that observation of nearly
 scale-invariant spectrum of curvature perturbation alone is
insufficient as a
 proof of inflation because other causal mechanisms can also exist to
 generate a similar spectrum \cite{Durrer:2001cg}.  
Such mechanisms, however, predict a positive correlation
 in TE and is now discriminated by WMAP observations \cite{Peiris:2003ff}.

Long wavelength components of the
stochastic gravitational wave background generated during inflation
can in principle be detected by the B-mode polarization of cosmic
microwave background radiation (CMB) \cite{Caldwell:1998aa},
and shorter-wave components with
frequency $\sim$ Hz may be observed by a space laser interferometer.
As for the former, although WMAP has obtained only an upper 
bound  \cite{Spergel:2003cb}, a
number of observational projects are  planned, and those include Planck
\cite{planck} and
BICEP \cite{bicep} suggesting a strong possibility for detection of
the stochastic background wave from inflation  by CMB.  On the other
hand,  
most theoretical  studies of the 
gravitational wave background of shorter wavelength 
have  emphasized on the detection
itself so far.  However a  discussion with an
anticipation for future 
diverse possibilities of gravitational wave cosmology, beyond the first
detection,  is worthwhile.

In the present paper,  we argue that the amplitude of high-frequency 
stochastic
gravitational wave background not only carries information on
the inflationary regime, during which they are generated, but also serves
as a probe of the equation of state in the early universe through which
the waves propagate towards us.  This double role played by
primordial gravitational wave background in  particle cosmology 
 is similar to that played by high-redshift quasars in
observational cosmology, which not only reflects the properties of
high-redshift universe, at their location, but also carry line-of-sight
information in their absorption spectra.  In this sense, just as high-redshift
quasars are regarded as  lighthouses of the distant universe, we may regard
the inflation-produced gravitational wave background as a lighthouse that
can shed light on the early universe physics.

As is well-known, equation of state of the early universe, just after $t=1$
sec and $T=1$ MeV, has been accurately probed by  big-bang
nucleosynthesis (BBN) calculations and a comparison of their yield with
observations \cite{Cyburt:2003fe}, 
though we do not have any sensible means to probe it prior to this
epoch.  This is why we only impose a mild  
constraint that the universe should
be dominated by radiation before BBN when we work on particle-physics
model building of the early universe involving lately decaying particles.
Thus, if we could observe the spectrum of gravitational waves from a 
source with a known property in the early
universe, we could obtain important information and constrain high
energy physics with which we describe the evolution of the universe.

We start with a brief review of properties of a  gravitational wave
generated 
quantum mechanically during inflation with the purpose of demonstrating
our notation.
We introduce tensor perturbations, $h_{ij}$, around a spatially flat
Robertson-Walker metric as 
\beq
ds^2=-dt^2+a^2(t)\left( {\delta _{ij}+2h_{ij}} \right)dx^idx^j,
\eeq
with $a$ being the scale factor.
Decomposing the tensor metric perturbation to  Fourier modes as
\beq
h_{ij}=\sqrt {8\pi G}\sum\limits_{A=+,\times } {\int {{{d^3k} \over
{(2\pi )^{{3 \mathord{\left/ {\vphantom {3 2}}
\right. \kern-\nulldelimiterspace} 2}}}}}}\varphi _k^A(t)e^{ikx}e_{ij}^A,
\eeq
we find that the two independent degrees of freedom $\varphi^{A}$ behave as
two massless minimally coupled scalar fields, where $e_{ij}^A$
represents polarization tensor with 
$e_{ij}^Ae_{}^{ijA'}=\delta ^{AA'}$ for $A,~A'=+,\times$.
Applying quantum field theory of a massless minimally coupled field in
de Sitter spacetime, we find that the Fourier modes are characterized by the
following vacuum correlation,
\beq
\left\langle {\varphi _k^A(t)\varphi _{k'}^{A'}(t)} \right\rangle ={{H^2}
\over {2k^3}}\delta ^3\left( {k-k'} \right)\delta ^{AA'},
\eeq
so that the amplitude per logarithmic frequency interval is given by
\beq
h_{F} ^2 (f ) \equiv 2\left\langle {h_{ij} h^{ij} (f )} 
\right\rangle  = 4 \times 8\pi G\left( {\frac{{H(\phi) }}{{2\pi }}} \right)^2  = \frac{8}{\pi }\left( {\frac{{H(\phi)}}{{M_{Pl} }}} \right)^2 ,
\eeq
where $M_{Pl}=G^{-1/2}$ denotes the Planck scale.
Here $H(\phi)$ denotes the Hubble parameter during inflation to be
evaluated when the mode with comoving frequency $f$ left the Hubble
radius during inflation.  It is expressed  as
\beq
H(\phi) = \sqrt{\frac{8\pi}3}\frac{V[\phi(f)]^{1/2}}{M_{Pl}},
\eeq
in terms of  the inflaton  potential $V[\phi]$, where
$\phi(f)$ denotes the value of the
inflaton field when the tensor mode $f$ crosses the horizon during the
inflationary epoch. 
%%%%%%%%%%%%%%%%%
%\section{Effect of the equation of the state}
%%%%%%%%%%%%%%%%5

In this paper, we use the present frequency $f$ to characterize the
comoving wavelength of gravitational waves, and the
scale factor $a$ for the cosmic time $t$. 
For the spectrum of gravitational wave background,  it is convenient to use a
dimensionless quantity $\Omega_{GW}(f,a)$ defined by
\beq
\Omega_{GW}(f,a)=\frac{\rho_{GW}(f,a)}{\rho_{cr}(a)},~~~
\rho_{GW}(f,a)\equiv
\frac{h^2(f,a)}{16\pi G}\lmk\frac{2\pi fa_0}{a}\rmk^2, \label{6}
\eeq
where $\rho_{GW}(f,a)$ is the energy density of gravitational waves per
unit logarithmic frequency interval around $f$, $a_0$ is the current
value of the scale factor, and 
$\rho_{cr}(a)$ is the critical
density of the universe. In the early universe we can  identify the critical
density  $\rho_{cr}(a)$
to the total density of the universe $\rho_{tot}(a)$, as  the spatial
curvature is negligible.  

In a  power-law background $a(t)\propto t^p$ with $p<1$, each Fourier
mode behaves as
\beq
 h(f,a)\propto
a(t)^{\frac{1-3p}{2p}}J_{\frac{3p-1}{2(1-p)}}\lmk\frac{p}{1-p}
\frac{k}{a(t)H(t)}\rmk,~~~k=2\pi fa_0, \label{7}
\eeq
with Bessel function $J_n(x)$. Such that the amplitude of gravitational
wave takes a constant value,
$h(f,a)=h_F(f)$, until a mode reenters the Hubble radius $H^{-1}$ at 
$a=2\pi fa_0/H \equiv a_{in}(f)$, where the normalized
energy density reads
\beq
 \Omega_{GW}(f,a_{in}(f))= \frac{4}{3\pi}\lmk\frac{H(\phi)}{M_{Pl}}
\rmk^2= \frac{32}9 \frac{V[\phi(f)]}{M_{Pl}^4}. \label{infnorm}
\eeq
\if
The temperature of the universe $T$ at the horizon crossing time of a
mode $f$ is roughly given by
\beq
f\sim 1\lmk\frac{T}{10^6{\rm GeV}}  \rmk {\rm Hz} .
\eeq
\fi

Evolution of the relative energy density of the tensor modes within the Hubble
horizon depends on the equation of state of the universe as we see below.
 From  the definition of $\Omega_{GW}$, we have
\beq
\frac{d\ln \Omega_{GW}(f,a)}{d\ln a}=\frac{d\ln \rho_{GW}(f,a)}{d\ln
a}-\frac{d\ln \rho_{tot}(f,a)}{d\ln a} \lab{ev1}.
\eeq
From eqs.\ (\ref{6}) and (\ref{7}), the 
tensor modes well  within the horizon behave simply as  radiation, 
\beq
\frac{d\ln \rho_{GW}(f,a)}{d\ln a}=-4 \lab{ev2},
\eeq
and evolution of  the total energy density is given by
\beq
\frac{d\ln \rho_{tot}(f,a)}{d\ln a}=-3\lkk 1+w(a)\rkk \lab{ev3},
\eeq
where $w$ is defined by $w\equiv P/\rho_{tot}$ with pressure of the
universe, $P$. 
From  eqs.\ (\ref{ev1}-\ref{ev3}) we have
\beq
\frac{d\ln \Omega_{GW}(f,a)}{d\ln a}=3w(a)-1,
\eeq
or
\beq
\ln\Omega_{GW}(f,a_0)=\int^{a_0}_{a_{in}(f)}\lkk 3w(a)-1 \rkk d\ln a+\ln
\Omega_{GW}(f,a_{in}(f)).\lab{lnomega}
\eeq
For our analysis it is convenient to use the following expression that
relates the present energy density $\Omega_{GW}(f,a_0)$ at different
frequencies $f_1$ and $f_2$,
\beq
\ln\Omega_{GW}(f_2,a_0)-\ln\Omega_{GW}(f_1,a_0)=
\int^{a_{in}(f_1)}_{a_{in}(f_2)}\lkk 3w(a)-1 \rkk d\ln a 
+\ln \frac{V[\phi(f_2)]}{V[\phi(f_1)]}\lab{kihon}.
\eeq
The first term represents the effect caused by deviation of  the
equation of state from  $w=1/3$ and the latter one shows the difference
of the intrinsic  amplitudes $\Omega_{GW}(f,a_{in}(f))$
generated at the inflationary epoch.  
Using the slow-roll equation of motion,
\beq
 3H\dot\phi =-V'[\phi],
\eeq
we can write down the latter term  as 
\beq
\ln \frac{V[\phi(f_2)]}{V[\phi(f_1)]}\simeq \frac{-M_{Pl}^2}{8\pi}\lmk\frac{V'}{V}
\rmk^2 \ln \lmk
\frac{f_2}{f_1}\rmk , \lab{correction}
\eeq
with $V'[\phi]\equiv dV[\phi]/d\phi$.
Here the coefficient ${M_{Pl}^2}/{8\pi}\lmk{V'}/{V}\rmk^2$ is
evaluated around $\phi\sim \phi(f_1)$  and  would be much smaller than
unity, which is one of the slow-roll parameters.

As is well known, under the assumption that fluctuations are governed by
a single inflaton field $\phi$, we can obtain the magnitude and the
first derivative of the inflaton potential $V[\phi]$ from the quadrupole
anisotropy of the cosmic microwave background radiation (CMB).  Denoting the
scalar and the tensor contributions by $S$ and $T$, respectively, we find
\beq
S_{l=2}=\frac{2.2 (V/M_{Pl}^4)}{(M_{Pl} V'/V)^2},~~~T_{l=2}=0.61 \lmk
\frac{V}{M_{Pl}^4}  \rmk, \lab{amp}
\eeq
where $V$ or $V'$ is evaluated at  $\phi(f)$ with mode $f$ corresponding to  the present
horizon $2\pi f=H_0$, and the values of the numerical coefficients are
those in the case of the Einstein de Sitter universe 
 \cite{Liddle:1994cr,Turner:1997ck}. 

As for the spectral indices $n_S$ and $n_T$ for these two modes, there
are deviations from the results ($n_S=1$ and $n_T=0$) for purely de-Sitter inflation due to
variation of the potential $V[\phi(f)]$ along with the modes  $f$  \cite{Liddle:1994cr,Turner:1997ck},
\beq
n_S-1=-\frac{M_{Pl}^2}{8\pi}\lmk\frac{V'}V
\rmk^2+\frac{M_{Pl}^2}{4\pi}{
\lmk\frac{V'}V
\rmk'}, ~~~n_T=\frac{M_{Pl}^2}{8\pi}\lmk\frac{V'}V
\rmk^2.\lab{spc}
\eeq
With eqs.\ (\ref{amp}) and (\ref{spc})  the coefficient
${M_{Pl}^2}/{8\pi}\lmk{V'}/{V}\rmk^2$ is expressed in a well-known form 
\beq
-\frac{M_{Pl}^2}{8\pi}\lmk\frac{V'}V
\rmk^2=n_T=-\frac{r}{7.0},~~~~r\equiv \frac{T_{l=2}}{S_{l=2}}. \lab{consist}
\eeq
However, the factor 7.0 in the above expression should be replaced by 5.0 for
realistic values of the cosmological parameters in the concordance model
\cite{Knox:1995dq}. 
WMAP has obtained an upper bound on the coefficient in 
eq.\ (\ref{correction}), as
$|{M_{Pl}^2}/{8\pi}\lmk{V'}/{V}\rmk^2|\sim r/5\lesssim0.1$ 
 \cite{Spergel:2003cb}. 
Thus even in the frequency difference of an order of magnitude,  $\ln
(f_2/f_1)\sim 2$, the second term of  eq.\ (\ref{kihon}) would be  very
small. If we could measure the difference of 
$\ln\Omega_{GW}(f_2,a_0)-\ln\Omega_{GW}(f_1,a_0)$,   for example to be
$\sim 
0.5$, it would  mainly be due to the first term. Namely, we could detect a
curious 
behavior of  equation of state $w\ne 1/3$ in  the  early universe.

As argued above,
there are essentially two frequency bands which may be regarded as a
serious target for observations.  
One is the ultra-low frequency band with
$f\sim H_0$, which can be detected indirectly through
B-mode polarization of CMB.
There are some CMB projects  aiming to measure the tensor fluctuations
as a  
primary target. 
As explained earlier,  BICEP, which is expected to start observations in
2004, 
 has a sensitivity down to $r\sim 0.1$ \cite{bicep}.
 CMBPOL, a space mission  planned to be launched
around 2012,  has sensitivity down to $r\sim 0.0001$ which is close to  the
fundamental limitation for the detection of $r$ using CMB polarization 
\cite{Knox:2002pe}.
 Planck, which is scheduled to be launched in 2007 \cite{planck},
 will determine the index $n_S$ with error less 
than $\Delta n_S<0.005$.  With these
future CMB projects,  we
will observationally obtain information related to the potential
$V[\phi]$ at a large 
scale,  $f\sim H_0$,  and  can use it to estimate  $V[\phi]$ up to the
second derivative using (\ref{spc}). 
Thus, although this frequency band is irrelevant to the evolution of the
cosmic equation of state, the information obtained through these
observations are invaluable that allow us to extrapolate
the spectrum $\Omega_{GW}(f,a_{in}(f))$ up to much higher frequencies 
 \cite{Turner:1997ck} which can be used as a calibrator in order to
 investigate an  anomalous change of the equation of state.

At present we have only an upper bound on tensor perturbations derived
from COBE and WMAP, namely, $H(\phi(f=H_0))< 2\times 10^{-5}M_{Pl}$ \cite{Spergel:2003cb}.
Without taking a possible change in $V[\phi]$ during inflation into
account, this corresponds to a constraint  $\Omega_{GW}(f\sim 1{\rm
Hz},a_0)< 8\times 10^{-16}$.  Thus WMAP data do not provide us direct
information on  the gravitational wave background.
If we consider  more
specific inflationary models, however,  we can predict a  precise value of
$\Omega_{GW}$ in the high-frequency region.  Among many mechanisms of
inflation, chaotic inflation proposed by Linde \cite{ci} 
is the most attractive for
its naturalness of initial condition.  WMAP, however, has ruled out one
of them, namely a model with a quartic potential
$V[\varphi]=\lambda\varphi^4/4$ \cite{Peiris:2003ff}.
The only remaining renormalizable model with a massive scalar potential,
on the other hand, has been shown to be not only in good agreement with
WMAP data \cite{ellis} but also well motivated by particle physics and
phenomenologically successful if we
identify the inflaton with a sneutrino field \cite{msyy}. 
This model predicts $r=0.16$ on large scales
 \cite{ellis}, which is within the reach of BICEP
\cite{bicep}.  
We can also 
calculate $\Omega_{GW}$ in the observable frequency range by laser
interferometers as $\Omega_{GW}(f\sim 1{\rm Hz}, a_0)=2.5\times
10^{-16}$.  Thus
allowing a room of uncertainty in inflationary models, it is desirable to
make a detector sensitive to $10^{-20}\lesssim \Omega_{GW} \lesssim 10^{-15}$.

The sensitivity of a laser interferometer 
is often represented by  the dimensionless gravitational 
wave amplitude $h_c(f)$.
For a detection of the  stochastic wave background,   correlation
analysis using multiple detectors is  very powerful 
\cite{Maggiore:2000vm,Allen}, and a detector with sensitivity
$h_c(f)$ can probe  $\Omega_{GW}$  up to
\beq
\Omega_{GW}(f,a_0)\simeq 10^{-16}X\lmk \frac{h_c(f)}{ 10^{-24}} \rmk^2
\lmk\frac{f}{1{\rm Hz}} \rmk^2\lmk\frac{\Delta f}{1{\rm Hz}}  \rmk^{-1/2}\lmk\frac{\Delta T}{1{\rm yr}}  \rmk^{-1/2}  \lab{omh},
\eeq
where $\Delta T$ is the observational period, $\Delta f$ is the effective
bandwidth, and the coefficient $X$ represents the overlap of two detectors
and is of order unity \cite{Maggiore:2000vm,Allen}. 
%Of course this sensitivity is that required for only detection of the
%background. We need somewhat better sensitivity to discern the frequency
%dependence of $\Omega_{GW}(f,a_0)$.
The initial LIGO has a  sensitivity corresponding to 
$\Omega_{GW}\sim 5\times
10^{-6}$ at $f\sim 100$Hz with $\Delta T=4$ months \cite{Allen}.
  The advanced LIGO
can detect  $\Omega_{GW}\sim 5\times
10^{-11}$ for the same observational period. As is apparent from
eq.\ (\ref{omh}), a lower frequency band is more advantageous for the
detection of gravitational waves with a nearly scale-invariant spectrum,
$\Omega_{GW}\simeq$ constant. LISA, which is scheduled 
to be launched in 2011,  has the best
sensitivity at $f \sim 10^{-3}$Hz \cite{lisa}. 
Although we cannot make  correlation
analysis with LISA due to the vanishing overlap of  data streams whose
noises do not correlate, we could attain the level of $\Omega_{GW}\gtrsim
10^{-12}$ \cite{Ungarelli:2001jp}. 
Unfortunately, however, the LISA band is severely contaminated
 by astrophysical sources and would hamper measurement of the primordial
background with $\Omega_{GW}\lesssim
10^{-11}$, because there are so 
many close white dwarf  binaries that they 
cannot be resolved  separately, which result in a confusion noise
 \cite{hils}.  Their contamination 
is expected to disappear  around $f\gtrsim 0.2$Hz due to  their large radii.
The frequency band $f\sim 1$Hz is expected to be
 relatively clean  \cite{Ungarelli:2001jp,decigo}, 
and this band could be observed by DECIGO \cite{decigo}.  If its
ultimately intended sensitivity is achieved, we could detect the
stochastic background up to $\Omega_{GW}\sim 10^{-20}$ at $\sim 0.2$Hz. 
Thus DECIGO would serve as an ideal detector to study the stochastic
gravitational wave background from inflation.

The comoving wave corresponding to $f\sim 1$Hz today reentered the 
Hubble horizon around $t\sim 3\times 10^{-20}$ sec with the cosmic
temperature $T\sim 3\times 10^6$GeV in the standard cosmology.
Thus, if significant entropy production takes place around
$t\sim 10^{-20}$ sec, we can detect its imprint on the spectrum of
$\Omega_{GW}$ using DECIGO.  

A typical mechanism of entropy production
in this epoch is decay of massive particles, which we denote by
$\chi$, with the decay rate $\Gamma_\chi
\sim 10^{-7}-10^{-5}$GeV.  If this particle decays through gravitational
interaction,  its mass corresponds to $m\sim (0.8-4)\times 10^{10}$ GeV.
Evolution of the energy
density of two components,  the radiation (relativistic particles) and the
particle $\chi$  is solved with the following equations
\beq
\frac{d\rho_{rad}(t)}{dt}=-4H\rho_{rad}(t)+\Gamma_\chi \rho_\chi(t),~~
\frac{d\rho_{\chi}(t)}{dt}=-3H\rho_{\chi}(t)-\Gamma_\chi \rho_\chi(t),
\eeq
where $\Gamma_\chi$ is the decay rate of the particle $\chi$ to
radiation. 
These equations are solved as
\beqa
\rho_{\chi}(a)&=&\rho_{\chi}(a_i)\lmk\frac{a}{a_i}\rmk^{-3}e^{-\Gamma_\chi
(t-t_i)},~~~a=a(t),\\
\rho_{rad}(a)&=&\rho_{rad}(a_i)\lmk\frac{a}{a_i}\rmk^{-4}
+\Gamma_\chi\lmk\frac{a}{a_i}\rmk^{-4}\int_{t_i}^{t}
\frac{a(\tau)}{a_i}\rho_{\chi}(a_i)e^{-\Gamma_\chi (\tau-t_i)}d\tau,
\eeqa
where $a_i=a(t_i)$ and $t_i \ll \Gamma_\chi^{-1}$ is an 
initial time with $\rho_{rad}(a_i)\gg \rho_\chi(a_i)$. 

Figure 1 depicts $\log\Omega_{GW}(f,a_0)$ in chaotic inflation model
driven by a massive scalar field with various values of $\Gamma_\chi$
and the entropy increase factor $F$.  The latter is defined by
\beq
  F\equiv \frac{s(t_e)a^3(t_e)}{s(t_i)a^3_i},~~~ s=\frac{4\pi^2}{90}g_\ast
T^3=\frac{4}{3}\frac{\rho_{rad}}{T},
\eeq
with $s$ being the entropy density, and $g_\ast$ is the effective number
of relativistic degrees of freedom. 
Here $t_e \gg \Gamma_\chi^{-1}$ is an arbitrary time in the radiation
dominated regime after entropy
production has terminated.  In this figure,  solid line represents the
case with $\rho_\chi=0$ from the beginning as a calibrator.  The weak
deviation from a flat straight line is due to the weak time dependence
of the Hubble parameter during inflation as described by the second term
in the right hand side of eq.\ (\ref{kihon}).  In the present model of
a massive scalar inflaton, this curve is represented as
$\Omega_{GW}(f,a_0)=2.5\times 10^{-16}\lkk 1-0.1\ln\lmk f/1{\rm
Hz}\rmk\rkk$. In this figure, dash-dotted line at the bottom depicts the
case with $\Gamma_\chi=10^{-7}$GeV with the entropy increase by a factor
$F=10^3$.  Three intermediate curves, on the other hand, all correspond to the
case $F=10$ but with different values of $\Gamma_\chi$.

As  seen there, the ratio of the values of $\Omega_{GW}$ in the left
and the right plateau depends  only on the entropy increase factor (apart
from the weak dependence of the solid line on $f$).
The reason is simple.  As the mode $f$ reenters the Hubble radius,
$\Omega_{GW}(f,a)$ evolves as eq.\ (\ref{ev1}) or
\beq
 \Omega_{GW}(f,a)=\frac{\rho_{tot}(a_{in})a_{in}^4}{\rho_{tot}(a)a^4}
 \Omega_{GW}(f,a_{in}(f)),  \label{eev}
\eeq
with the initial condition given in eq.(\ref{infnorm}).  
Let us consider $\Omega_{GW}(f,a)$ at $a=a_e=a(t_e)$ with
$t_{in}(f)<t_e$.  If $t_{in}(f)\gg \Gamma_{\chi}^{-1}$ the mode reenters the
horizon after the entropy production has terminated, so the factor in
the right hand side of eq.\ (\ref{eev}) is equal to unity and we find
$\Omega_{GW}(f,a_e)=\Omega_{GW}(f,a_{in}(f))$ in this case, which
corresponds to the left plateau in  figure 1.  For the modes that
reenter the horizon sufficiently early, we have $\rho_{tot}(a_{in})a_{in}^4
=\rho_{rad}(a_{in})a_{in}^4$, so we find
$\Omega_{GW}(f,a_e)=F^{-4/3}\Omega_{GW}(f,a_{in}(f))$, because
$\rho_{tot}$ is also equal to $\rho_{rad}$ at $a=a_e$.   
Here we have assumed that $g_\ast$
does not change between $t_{in}(f)$ and $t_e$.  This region corresponds
to the right plateau in the figure.  Thus the ratio is given by
$F^{-4/3}$ apart from the weak dependence on $V[\phi(f)]$.

If the entropy production occurs at a later epoch with $\Gamma_\chi \ll
10^{-7}$ GeV, the entire frequency band observable with DECIGO lies in
the right plateau region in the figure.  We could still estimate the
entropy increase factor $F$ in such a case if we could measure the
Hubble parameter during inflation with CMB polarization and extrapolate
it to a higher frequency band.

So far we have  assumed that inflation-produced tensor
perturbations are  the only source of the stochastic gravitational wave
background apart from possible contamination from  Galactic and
extra-Galactic  stellar binaries. 
But, there are several proposed cosmological sources of
backgrounds, such as cosmic strings \cite{VS}
 and field rearrangement associated
with a global phase transition \cite{Krauss:1991qu}.  
The former produces
\beq
\Omega_{GW}(1{\rm Hz}, a_0)=3\times 10^{-8}\lmk\frac{v_L}{10^{16}{\rm
GeV}}\rmk,  \label{string}
\eeq
where $v_L$ is the symmetry breaking scale of string formation 
\cite{Caldwell:1996en}.  This would exceed inflationary gravitational
radiation if $v_L>10^{8}$ GeV.  On the other hand, global phase transition
induces 
\beq
 \Omega_{GW}(f,a_{in}(f))\simeq \frac{8\pi}3 \frac{v_G^4}{M_{Pl}^4},
 \label{phasetr} 
\eeq 
which should be compared with eq.\ (\ref{infnorm}) \cite{Krauss:1991qu}.  
Thus, in order to suppress the contribution related to the symmetry
breaking scale,  
$v_G$ should be smaller than $H(\phi)$ during inflation.
Even if these contributions surpass that of inflation, however, we can probe
equation of state in the early universe with these gravitational waves
because they also have intrinsically scale-invariant spectrum at the
outset.
(Note that eqs.\ (\ref{string}) and (\ref{phasetr}) apply
to the case phase transitions occur after inflation.  If the symmetry
breaking scales are so large that phase transitions take place in the
early stage of inflation, they do not contribute to the stochastic
gravitational wave background.  Alternatively, if phase transitions
occur
 in a late stage of inflation \cite{nonminimal}, 
they might leave an interesting
imprint on the stochastic background observable with laser interferometers.)

In summary we have proposed to use the stochastic
gravitational wave background from  inflation to probe  the
equation of  
state of   early universe and its evolution. Due to the  nearly scale
invariant nature of 
inflation, small deviations from a simple radiation dominated universe
($w=1/3$) 
would be clearly imprinted in the spectrum of  $\Omega_{GW}(f,a_0)$. We
could obtain interesting information that can be hardly extracted by other
methods. 

\vskip 1cm

N.S. would like to thank A. Cooray for useful discussions and JSPS for a
fellowship to research abroad.
This work was supported in part by JSPS
Grant-in-Aid of Scientific Research  No.\ 13640285(JY).

\newpage

\begin{figure}[h]
 \begin{center}
 \epsfxsize=14.cm
 \begin{minipage}{\epsfxsize} \epsffile{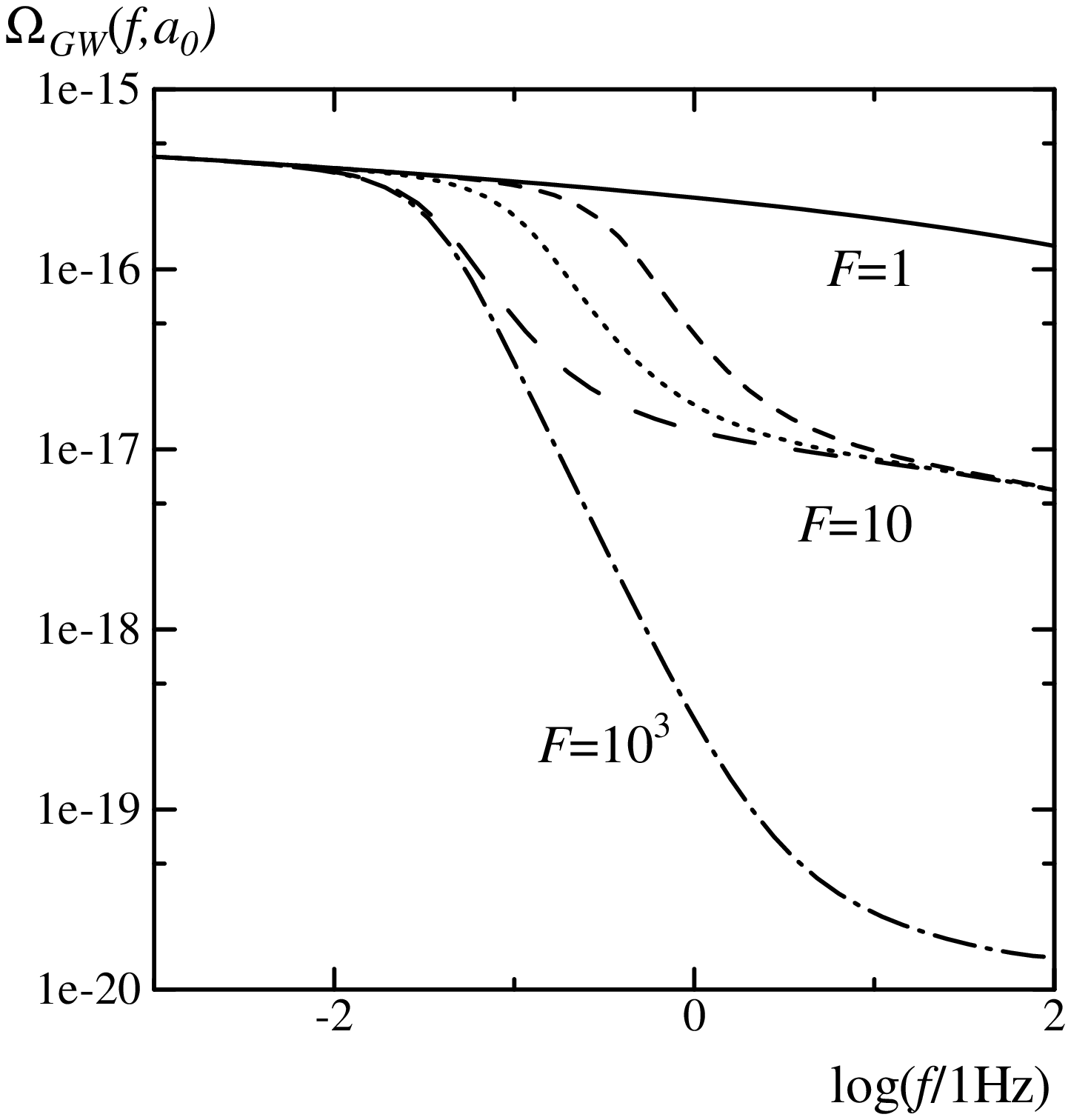} \end{minipage}
 \end{center}
\caption[]{ Energy  spectrum of the gravitational wave background, 
$\Omega_{GW}(f,a_0)$, in a  chaotic inflationary model driven by a massive
 scalar field.  Solid line represents the case with no entropy
 production. Dash-dotted line depicts the case with
 $\Gamma_\chi=10^{-7}$ GeV and the entropy increase factor $F=10^3$.
Three intermediate curves are for $F=10$ with $\Gamma_\chi=10^{-5}$ GeV
 (short-dashed line), $\Gamma_\chi=10^{-6}$ GeV (dotted line),
and  $\Gamma_\chi=10^{-7}$ GeV (long-dashed line), respectively.} 
\end{figure}

\end{document}